# Leveraging Contextual Data Augmentation for Generalizable Melanoma Detection

Nick DiSanto, Gavin Harding, Ethan Martinez, Benjamin Sanders
California Baptist University
Email: {nicolasc.disanto, gavinjames.harding, ethandaniel.martinez, bsanders}@calbaptist.edu

*Abstract* - **While skin cancer detection has been a valuable deep learning application for years, its evaluation has often neglected the context in which testing images are assessed. Traditional melanoma classifiers assume that their testing environments are comparable to the structured images they are trained on. This paper challenges this notion and argues that mole size, a critical attribute in professional dermatology, can be misleading in automated melanoma detection. While malignant melanomas tend to be larger than benign melanomas, relying solely on size can be unreliable and even harmful when contextual scaling of images is not possible. To address this issue, this implementation proposes a custom model that performs various data augmentation procedures to prevent overfitting to incorrect parameters and simulate real-world usage of melanoma detection applications. Multiple custom models employing different forms of data augmentation are implemented to highlight the most significant features of mole classifiers. These implementations emphasize the importance of considering user unpredictability when deploying such applications. The caution required when manually modifying data is acknowledged, as it can result in data loss and biased conclusions. Additionally, the significance of data augmentation in both the dermatology and deep learning communities is considered.**

## 1 Introduction

Melanoma, a highly lethal form of skin cancer, has become increasingly dangerous in the early twenty-first century. With millions of people across the globe being diagnosed with skin cancer every year [1], the incidence of melanoma has risen by over 250% over the last four decades, affecting one in sixty-three Americans [2]. Even while skin cancer treatment has grown increasingly sophisticated and the incidence of many other tumors is decreasing, melanoma has remained a major cause of mortality, claiming nearly ten thousand lives annually in the United States alone [3]. Early detection is widely recognized as the key to lowering the melanoma mortality rate, as the disease can become life-threatening in as few as six weeks [4, 5]. Licensed dermatology, the traditional route to accurate diagnosis, relies on visual inspection and the description of particular lesions for accurate diagnosis [6]. However, professional treatment is often inconvenient, time-consuming, and costly for the average individual. As medical facilities are often overwhelmed with patients, the development of automated detection processes has become a crucial priority. Machine learning approaches, including deep learning solutions like Convolutional Neural Networks (CNNs), have shown promising results in melanoma detection and classification. Unfortunately, these algorithms often fall short in clinical settings, with some experiencing accuracy drop-offs of up to 29% [7].

In order to develop CNNs that can accurately identify individual images of melanoma, it is essential to teach them the most significant features of skin cancer classification. According to leading dermatologists, a mole's three most important characteristics are its size, shape, and color [8, 9]. While these features may be straightforward for dermatologists to assess in person, it is vital to consider them in the context of a single image. If an image classification model can adequately understand each attribute within the context of a single image, a strong case can be made for dermatologist-level classification ability.

While CNNs show incredible potential in image classification, an important consideration is the context in which images are encountered. Images consistently high in quality and uniform in execution are more likely to yield deep and robust neural networks. However, unregulated testing images make accurate classification less predictable. Riberio [10] demonstrates a fitting example, accidentally creating a model which classified images based on the appearance of snow instead of learning the distinction between dogs and wolves. The risks associated with these hidden variables are critical to consider in the case of skin cancer. In order to optimize a melanoma classifier, it is essential to understand whether the model's attributes depend on the nature or quality of its images. If external factors impact the model's prediction, data augmentation techniques must ensure proper capturing and understanding of the image's relevant characteristics.

When isolated and trained on, the shape of a cancerous mole is certainly the most straightforward attribute for a deep learning model to understand. With the exception of incredibly low-resolution images, the shape of a lesion is nearly impossible to augment artificially, making generalization abilities much more robust. Due to this simplicity, many skin classifiers preprocess their data to specifically highlight mole shape [11]. Similarly, a mole's color is a feature that CNNs have a high potential to substantially understand [12]. The color of a photo provides significant insight to a mole classifier regardless of the photo's quality. After all, color remains discernible regardless of whether an image is of low or high resolution. However, as Kandel [13] points out, a necessary consideration for color is the potential inconsistency in image brightness and contrast in individual images.

The final and most critical attribute to detecting a cancerous mole is its diameter. Malignant melanomas tend to be larger than 6mm and exhibit rapid growth, while benign lesions can be up to 20% smaller and do not spread or enlarge [14]. This distinction is usually dermatologists' quickest and most definitive consideration when visually assessing a lesion. However, a mole's size is notably unpredictable. It is often unclear how closely an image is being taken, forcing networks to compare photos of varying zoom as if they are analogous in inherent structure. To meet the needs of medical diagnoses, algorithms must not necessarily conflate the size of a mole relative to the photo with its actual diameter. Given this uncertainty, this paper argues that training an image classifier on data that emphasizes the significance of mole size can harm the network's performance in situations where testing images may not be regulated. The goal is to demonstrate the importance of aligning a model's specific features and its respective understanding of them.

## 2 RELATED WORKS

While medical imaging has witnessed widespread utilization of CNNs and other deep learning architectures since 2012 [15, 16], the recent soar in popularity of skin cancer classification can be primarily attributed to the annual ISIC Challenge [17]. The International Skin Imaging Collaboration (ISIC), which houses the world's most extensive collection of publicly available skin lesion datasets, has been sponsoring challenges since 2016 for the computer science community to produce algorithms that can outperform professional dermatologists [18]. This competition has sparked the development of complex algorithms and yielded impressive results in melanoma classification [19, 20], significantly advancing the goal of automated melanoma detection and contributing to improvements in state-of-the-art deep learning approaches [21, 22]. This paper will use the 2018 ISIC dataset for melanoma classification.

CNNs are certainly the most common algorithms for skin cancer detection, but other applications, such as Support Vector Machines, k-Nearest Neighbors, and decision trees, have also produced impressive results [23]. Each approach offers distinct ways to extract and learn from features present in the training data. Javaid [24] discusses the important distinctions among these methods, providing insights into their optimized usage scenarios. Furthermore, within the realm of deep learning, various models such as ResNet, Xception, VGG19, and AlexNet have shown promising results [25, 26]. While every approach emphasizes different aspects of the skin cancer classification problem, they each demonstrate valuable and unique representations of the data.

CNNs are often susceptible to overfitting, and skin cancer classification provides no exception. Maron [27] illustrates examples of noise, such as skin markings, that can pose challenges to optimization. Similarly, Hekler [28] highlights the profound impact even minimal label noise can have on network accuracy. These factors are crucial considerations when aiming to achieve dermatologist-level performance, as they present significant hurdles in automated systems that human professionals can avoid with simple intuition. Many previous skin cancer classifiers have employed various image augmentation techniques to mitigate overfitting. For instance, Rezaoana [29] applies image shearing and horizontal flipping, while Junayed [30] incorporates randomized flipping and shading. These experiments serve as sound examples of methods that enhance the generalization capabilities of CNNs. Additionally, data preprocessing techniques, such as normalization, can be highly beneficial when performing transfer learning on deep CNNs [31].

Numerous published manuscripts extol the benefits of resizing melanoma images to address overfitting concerns. Tabrizchi [32] demonstrates significant improvements in VGG models through image resizing. Furthermore, Mikołajczyk & Grochowski [33] employ affine transformations on uneven training data to assess the effectiveness of each method. While these demonstrations certainly improve accuracy extreme care is required to avoid manipulating training data. Nunnari & Sonntag [34] highlight the risks associated with excessive augmentation, revealing potential data loss during haphazard resizing. This approach aims to validate the advantages of data augmentation while adopting a conservative approach to mitigate the risk of data loss.

## 3 EXPERIMENTS

*3.1 – Dataset*

This study utilizes the ISIC Archive's dataset, which is comprised of dermoscopic images of malignant and benign skin lesions, as shown in Figure 1 [35]. These images are sponsored by the International Society for Digital Imaging of the Skin (ISDIS), an international organization that publicly provides quality-controlled images of skin lesions. It is important to note that these images undergo a vetting process by melanoma experts who help to review and classify them, ensuring the data's quality and accuracy. The dataset used in this study consists of 1686 training images, along with 213 testing images and 210 validation images. To align with the initialization of the pre-trained layers, every image is initially normalized to a size of 224x224 pixels.

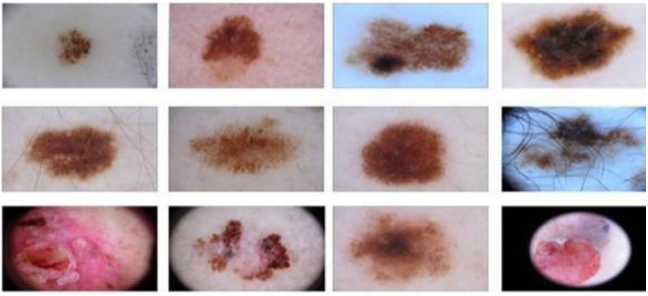

FIGURE I
IMAGES FROM ISIC MELANOMA DATASET

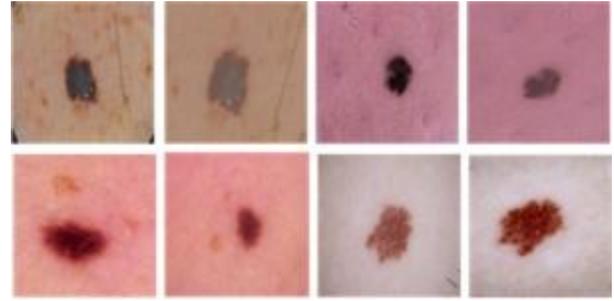

FIGURE II
EXAMPLES OF ROTATION, CONTRAST, AND ZOOM AUGMENTATION

*3.2 – Preprocessing Techniques*

Data preprocessing techniques are necessary to simulate arbitrary image capturing in the testing data. Various data augmentation techniques are used to analyze the effectiveness of the model's generalization abilities. One such technique involves randomly rotating and adjusting the contrast of each image by up to 50%, as demonstrated in Tables I and II. This augmentation provides a realistic representation of the varying image qualities associated with each mole.

The other preprocessing method employed is resizing, as similarly demonstrated below. This technique evaluates the effectiveness of eliminating mole size. In order to neutralize the size difference between malignant and benign melanomas, each image of benign melanoma is randomly zoomed in up to 10%, while each image of malignant melanoma is randomly zoomed out up to 10%. This process effectively removes size as a statistically relevant feature and ensures that the model's training focuses only on the appropriate characteristics. Additionally, one model was given images that were inversely resized to verify that its overfitting would increase. After augmenting each model's training data to generalize its usefulness, the testing data is also augmented by up to 50% to simulate arbitrary use.

| Training Augmentation Range | | |
|---|---|---|
| Method | Benign | Malignant |
| Rotation | {-50% – 50%} | {-50% – 50%} |
| Contrast | {-50% – 50%} | {-50% – 50%} |
| Zoom | {0% – 10%} | {-10% – 0%} |
| Inverse Zoom | {-10% – 0%} | {0% – 10%} |

TABLE I
AUGMENTATION METHODS RANDOMLY PERFORMED ON TRAINING DATA

| Testing Augmentation Range | | |
|---|---|---|
| Method | Benign | Malignant |
| Rotation | {-50% – 50%} | {-50% – 50%} |
| Contrast | {-50% – 50%} | {-50% – 50%} |
| Zoom | {-50% – 50%} | {-50% – 50%} |

TABLE II
AUGMENTATION METHODS RANDOMLY PERFORMED ON TESTING DATA

*3.3 – Custom VGG19 Model*

This implementation creates a custom model based on Keras' VGG19 Deep Convolutional Neural Network [36]. Multiple instances of the network are utilized so that they can be trained on different sets of augmented images. Initially, the models are trained on the ImageNet database, expediting the rudimentary training process. The following table represents the architecture of the model:

| Layer Type | Output Shape | Parameters |
|---|---|---|
| Input | 224, 224, 3 | 0 |
| Conv2D | 224, 224, 64 | 1792 |
| Conv2D | 224, 224, 64 | 36928 |
| MaxPooling2D | 112, 112, 64 | 0 |
| Conv2D | 112, 112, 128 | 73856 |
| Conv2D | 112, 112, 128 | 147584 |
| MaxPooling2D | 56, 56, 128 | 0 |
| Conv2D | 56, 56, 256 | 295168 |
| Conv2D | 56, 56, 256 | 590080 |
| Conv2D | 56, 56, 256 | 590080 |
| Conv2D | 56, 56, 256 | 590080 |
| MaxPooling2D | 28, 28, 256 | 0 |
| Conv2D | 28, 28, 512 | 1180160 |
| Conv2D | 28, 28, 512 | 2359808 |
| Conv2D | 28, 28, 512 | 2359808 |
| Conv2D | 28, 28, 512 | 2359808 |
| MaxPooling2D | 14, 14, 512 | 0 |
| Conv2D | 14, 14, 512 | 2359808 |
| Conv2D | 14, 14, 512 | 2359808 |
| Conv2D | 14, 14, 512 | 2359808 |
| Conv2D | 14, 14, 512 | 2359808 |
| MaxPooling2D | 7, 7, 512 | 0 |
| Flatten | 25088 | 0 |
| Dense (ReLU) | 256 | 6422784 |
| Dropout | 256 | 0 |
| Dense (ReLU) | 128 | 32896 |
| Dropout | 128 | 0 |
| Dense (ReLU) | 64 | 8256 |
| Dropout | 64 | 0 |
| Dense (SoftMax) | 2 | 130 |

TABLE III
LAYER REPRESENTATION OF PROPOSED MODEL

After initializing the networks and training on their respective datasets, transfer learning is applied to fine-tune each model's hyperparameters. This is simply with a goal to improve generalization to the testing data. Transfer learning enables us to leverage the pre-trained layers from ImageNet, while the subsequent layers are tailored to the specific dataset. Finally, the convolutional layers of each model converge into two-layer classifiers, producing the final prediction of the model: malignant or benign.

## 4 RESULTS

In order to investigate the significance of specific features in automated melanoma detection, several experiments are conducted using varying levels of augmentation. The accuracy of each model provides insights into the importance of each image component. Initially, the unmodified ISIC dataset was used for training, and the augmented testing images were evaluated. As anticipated, the model exhibited severe overfitting and, after twenty epochs of training, struggled to accurately classify the modified testing images. However, the first preprocessing techniques showed notable improvement. Consistent with previous studies [26], augmenting the training data with rotation and contrast dropped the generalization gap substantially.

Furthermore, the third model, which simply normalized size in the training images, resulted in an even greater jump in testing accuracy. This observation alone highlights the advantage of removing mole size as a distinguishing feature. Additionally, to promote generalization in terms of rotation, contrast, and zoom, the fourth network was trained with full augmentation: rotation, contrast, and resizing. This comprehensive approach resulted in the highest classification accuracy achieved thus far, effectively showing no drop in accuracy between training and testing environments.

To further establish the significance of the overfit models, the final model was trained on data resized in the opposite manner, enlarging malignant melanomas and shrinking benign melanomas. Expectedly, the inverse resizing of training images led to a decrease in accuracy compared to the untouched ISIC dataset, illustrating the potential consequences of an expected size disparity between malignant and benign moles. The complete results of these experiments are summarized below:

| Training Augmentation | Training Accuracy | Testing Accuracy | Training/Testing Difference |
|---|---|---|---|
| None | 84% | 67% | 17% |
| Rotation & Contrast | 80% | 73% | 7% |
| Resizing | 79% | 76% | 3% |
| Rotation, Contrast, & Resizing | 85% | 84% | 1% |
| Inverse Resizing | 83% | 63% | 20% |

TABLE IV
ACCURACIES BASED ON VARYING AUGMENTATION METHODS

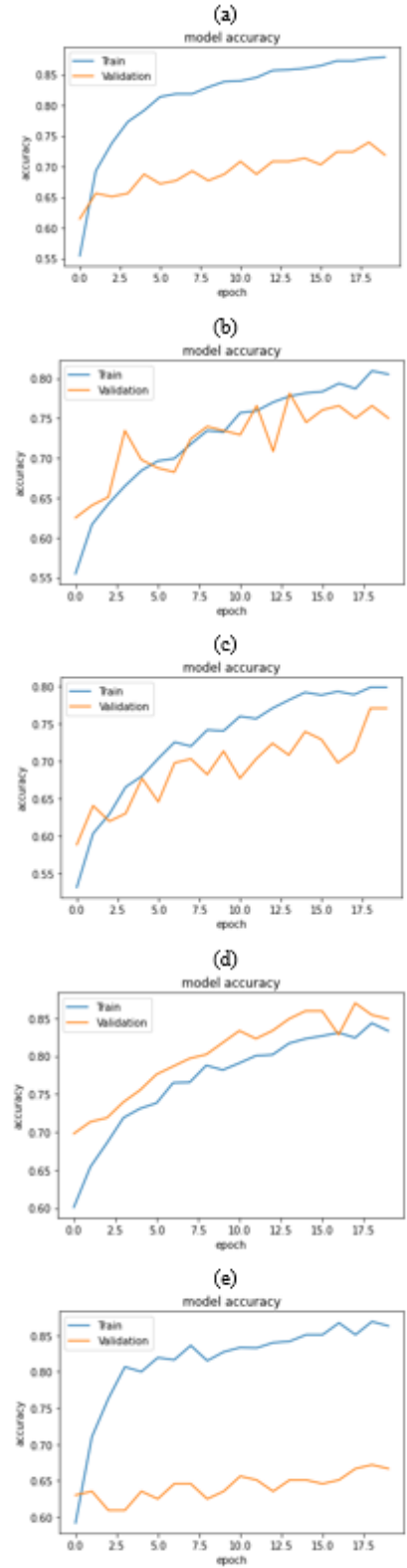

FIGURE III
TRAINING AND VALIDATION ACCURACIES AFTER EACH TRAINING AUGMENTATION METHOD: (A) NONE, (B) ROTATION & CONTRAST, (C) RESIZING, (D) ROTATION, CONTRAST, & RESIZING, (E) INVERSE RESIZING

Transfer learning provided moderate improvements to each of the models, with the exception of the overfitted model with opposite resizing augmentation. The large generalization gaps in the overfit models indicate a limited understanding of the data and a poor understanding of significant features. From these findings, two main conclusions can be drawn: image preprocessing can be highly effective in optimizing CNN performance in dermatology, and mole size can be a highly unreliable feature for testing deep learning models.

## 5 DISCUSSION

The results of this study highlight the importance of data augmentation and the consideration of mole size in melanoma classification models. The models that eliminated mole size as a feature showed improved accuracy, emphasizing the significance of accounting for user inconsistency. In real-world image classification applications, it is essential to acknowledge the variability in testing data, which poses a risk of overfitting when training data is organized too strictly. This should be considered when developing and evaluating algorithms in this field.

There is a tradeoff between accuracy and simplicity when comparing deep learning models with human dermatologists, as argued by Saarela and Geogieva [36]. The more accurate a model is, the more complex it tends to become, often with black-box hidden layers that are challenging to interpret. This is an important consideration when comparing neural networks and human dermatologists. Unfortunately, in order to develop a model which rivals or exceeds human-like classification ability, it seems inevitable that the model's complexity becomes unintuitive to the average bystander. However, making simple adjustments to preprocess data can improve a model's effectiveness in a real-world setting without increasing its underlying complexity.

Although this implementation demonstrates improved accuracy when preprocessing each model's training data, it is necessary to acknowledge the care required when manually modifying data in this manner. The volatility of skin detection models is evident in the variance of the results. If the sensitivity of melanoma image details is not considered, there is a significant risk of data loss, fundamentally altering the nature of the original problem. Balderas [37] distinguishes between beneficial preprocessing and biased modification. Data preprocessing is a fine line to tread, as slight alterations can drastically bias the data's representation [38].

While the model functions with higher accuracy when mole size is removed as a feature, it is also necessary to recognize the distinction that must be made between the model and a human dermatologist. While the model achieves high accuracy without considering mole size, it misses out on its potential benefits as a definitive variable. If size could be considered in the context of a single image, it would undoubtedly be advantageous to include as a consideration. Unfortunately, current machine learning interpretations do not easily lend themselves to such an advancement. Similarly, evaluating multiple images of a single mole and extrapolating a three-dimensional model of varying angles could provide additional information about the mole's relative diameter. While a method like this extends beyond the scope of this paper, it is a possibility to consider when innovating new methods of melanoma detection models.

## 6 CONCLUSION

This paper demonstrates a simple yet robust preprocessing approach for optimizing the classification of melanoma images. In particular, it draws attention to the potential risks of relying heavily on size as a substantial feature of mole classification. The importance of user unreliability is an attribute that cannot be ignored, and extreme intentionality is required to account for such inconsistencies. While many models can achieve high training accuracy, generalization abilities are essential for successful application in real-world scenarios. Applying these preprocessing techniques proves to be an effective means to diminish the role that size plays in mole classification. However, it is important to acknowledge that by eliminating this feature, a potentially critical distinction used by dermatologists to identify melanoma may be sacrificed. Therefore, the medical imaging community should focus resources on finding innovative ways to reintroduce size as a reliable feature to training data without compromising the model's effectiveness. Until machine learning models can better perceive external variables in their domains, they must account for limitations in the data they are trained on.